\documentclass[useAMS,usenatbib]{mn2e}
\usepackage{graphicx}
\usepackage{multirow}
\usepackage{amsmath}
\usepackage{longtable}

\title[]{\textit{STEREO/HI} and Optical Observations of the Classical Nova V5583 Sagittarii}

\author[Holdsworth et al.]
{Daniel. L. Holdsworth$^{1,2}$, M. T. Rushton$^1$, D. Bewsher$^1$, F. M. Walter$^3$, S. P. S. Eyres$^1$, \newauthor
R. Hounsell$^4$ and M. J. Darnley$^5$
\\
$^1$ Jeremiah Horrocks Institute, University of Central Lancashire, Preston, PR1 2HE. \\
$^2$ Astrophysics Group, Keele University, Staffordshire, ST5 5BG.\\
$^3$ Department of Physics \& Astronomy, Stony Brook University, Stony Brook, NY 11794-3800, USA.\\
$^4$ Space Science Institute, 3700 San Martin Dr, Baltimore, MD 21218, USA.\\
$^5$ Astrophysics Research Institute, Liverpool John Moores University, Liverpool, L3 5RF. 
}

\date{Version of September 2013}

\pagerange{\pageref{firstpage}--\pageref{lastpage}} \pubyear{2013}

\def\LaTeX{L\kern-.36em\raise.3ex\hbox{a}\kern-.15em
    T\kern-.1667em\lower.7ex\hbox{E}\kern-.125emX}
\begin{document}
\label{firstpage}
\maketitle

\begin{abstract}

The classical nova V5583 Sgr (Nova Sagittarii 2009 N$^o~3$) has been observed during the rise phase and shortly after by NASA's \textit{STEREO/HI} instruments, with later optical spectroscopy obtained with the R-C Spectrograph at CTIO, Chile. The time of peak in the STEREO passband has been constrained to within 4 hours, as a result of the high cadence data obtained by \textit{STEREO/HI}. The optical spectra show the nova evolving from the permitted to the nebular phases. The neon abundance in the ejecta is $\rm [Ne/O]\ga+1.0$, which suggests that V5583~Sgr was most likely a neon nova.

\end{abstract} 

\vskip2mm

\section{Introduction}

Classical nova (CN) eruptions occur in close binaries containing a white dwarf (WD) primary and a low mass star \citep[see][for recent reviews]{bode08}. In these systems, the secondary fills its Roche lobe and transfers hydrogen-rich material to the WD via an accretion disk. The outbursts occur when the mass accreted on the white dwarf reaches a critical value. Novae generate $\sim10^{37} - 10^{39}$ J and eject $\sim10^{-4} M_{\sun}$ into the Interstellar Medium (see \citealt{gehrz98} for a review). 

The Galactic nova rate has been estimated to be $35\pm11$ yr$^{-1}$ \citep{shafter97}, and $34^{+15}_{-12}$\,yr$^{-1}$ \citep{darnley06}. However, $<10$\,yr$^{-1}$ are well observed, because of our position within the galaxy and the interstellar extinction in the Galactic disc \citep{shafter97}. It is thought that a further $\sim5$ nova a year are missed because of the close proximity of the Sun on the sky \citep{hounsell10}.

Novae are usually discovered by amateurs and reported to the American Association of Variable Star Observers (AAVSO) or Variable Star NETwork (VSNET) databases. This means that novae tend to be discovered when they are brightest. The rise phase is rarely well observed, although it is known that at least some novae pause about 1-2 magnitudes below the peak, a phenomenon known as the pre-maximum halt (PMH; \citealt{Warner08}). 

\citet{hounsell10} presented light curves of four nova eruptions from an unlikely source: the Solar Mass Ejection Imager (SMEI) on board the \textit{Coriolis} satellite \citep{eyles03}, an instrument designed to observe coronal mass ejections (CMEs). Because of the continuous coverage of the SMEI data, they detected the novae before the peak, and tracked them into the decline. Due to the high cadence of the observations (102 minutes), they were able to construct nova light curves at unprecedented temporal resolution. PMHs were detected in three of the novae. 

The Heliospheric Imagers \citep[HI;][]{hi09} on board the Solar TErrestrial RElations Observatory \citep[\textit{STEREO};][]{kaiser2008} also have the potential to detect nova outbursts. They monitor the heliosphere for CMEs, observing the background stars. 
This data set, spanning nearly seven years, has recently been exploited for stellar work, by e.g. \citet{wraight11}, \citet{wraight12}.

V5583~Sagittarius (Nova Sagittarii 2009 N$^o$ 3) was discovered in eruption by \citet{IAUC9061} on 2009 August 6.494 UT at a magnitude of $V=7.7$. Optical spectroscopy showed broad Balmer and Fe\,{\sc ii} emission lines as expected for a classical nova \citep{Maehara09}. This interpretation was supported by the observations of \cite{v5583}, who found that the multi-band light curves were almost identical to those  of the classical nova V382~Vel. In this paper, we present the light curve for V5583~Sgr, the first nova detected in the HI instruments.

\section{Photometric Observations}

\subsection{STEREO/HI light curve}

\begin{figure*}
  \centering
  \includegraphics[width=160mm]{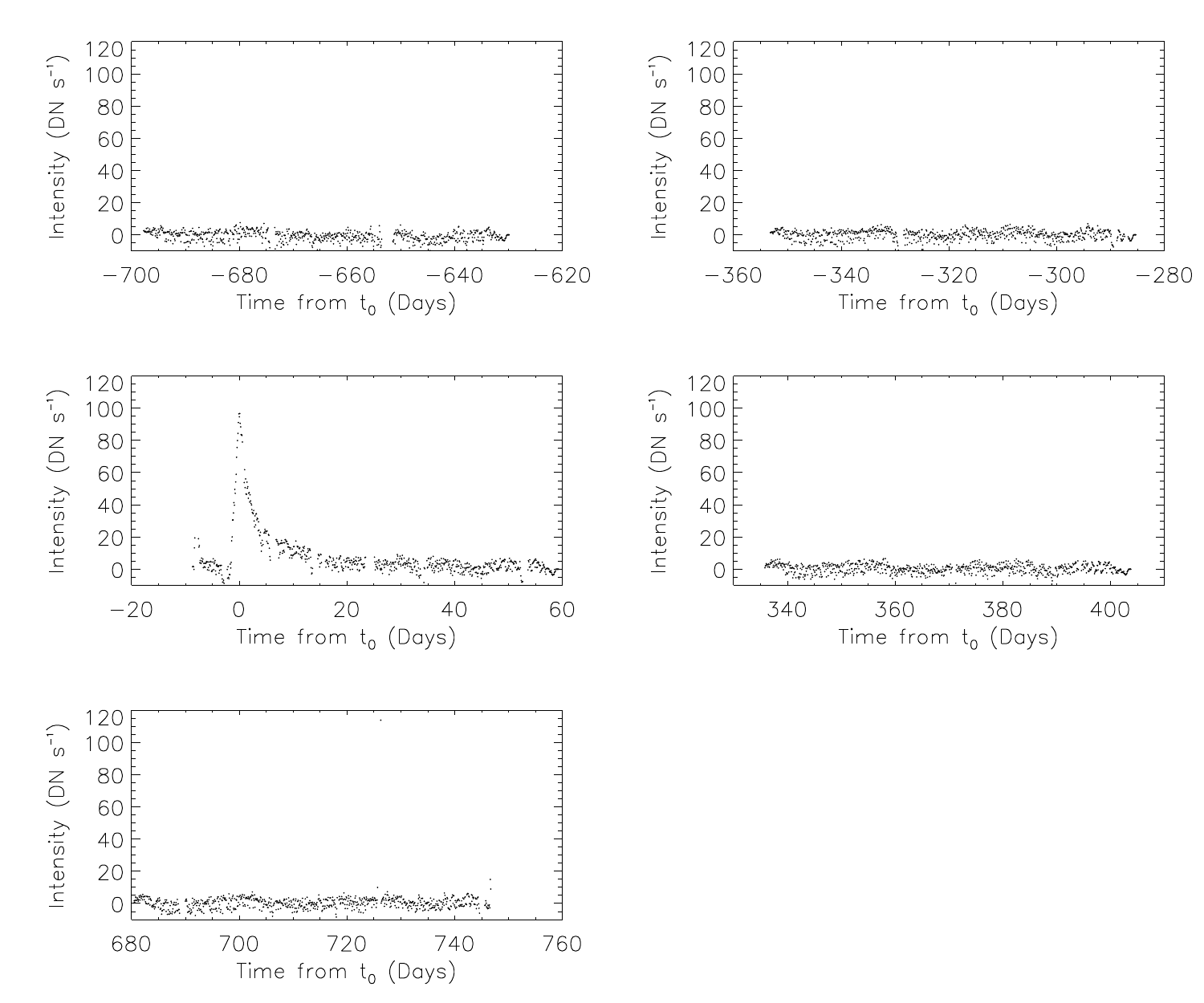}
  \caption{HI magnitude light curves of V5583 Sgr over the first 5 years of HI-2A data. The outburst can clearly be seen in the 3rd epoch of data.}
  \label{fig:hi_lc}
\end{figure*}

The HI observations used in this paper were obtained by the Heliospheric Imagers (HI-2) on the STEREO-Ahead (STEREO-A) spacecraft.

The HI-2 images have a field of view (FOV) of $\sim 70^\circ \times 70^\circ$ with a pixel size of $\sim 4$ arcmin pixel$^{-1}$, centred $53^\circ$ along the ecliptic plane from Sun centre. The HI-2 cameras have a broad spectral response ($\sim4000-10000$\AA) with the aim of detecting as much white light from faint CMEs as possible \citep{hi09}. Niney-nine exposures of 50 seconds each, make up each image, with an image cadence of 120 minutes. The solar F-corona is the dominant large scale structure in the images, which is removed using a running daily background.

V5583~Sgr entered the FOV of the HI-2A camera on 2009 July 29 UT (8 days before its discovery), and left the FOV on 2009 October 5 UT. 
As this is the first stellar observation using data from HI-2A, we give a short overview of the data analysis.

The HI-2 instruments have baffles which prevent stray-light from entering the optical system, but result in a reduced portion of the FOV being used for science purposes. Therefore, in this work, we exclude data in the baffled regions. 

At the time of writing, a post-launch calibrated largescale flatfield is unavailable for the HI-2A instrument. However, two stars located on the same part of the CCD as the nova have been used to estimate the large scale flatfield using the same methodology that \citet{bewsher10} used for the HI-1 instruments. This correction for the flatfield has been applied to the data. 

In each HI-2A image, the position of the nova is determined using the pointing and optical calibration of \citet{brown2009}. An aperture with radius of 2.18 pixels is used to calculate the intensity of the nova. This is repeated on every HI-2A image which contains the nova location over a 5 year period, which corresponds to crossing the CCD five times. The nova outburst is observed during the third of the five crossings of the CCD. The aperture size was chosen so that the aperture was large enough to render any PSF variation negligable, yet avoid any contamination from nearby sources. Catalogues show a number of other sources within the aperture, but all are faint (V $> 10$) and below the sensitivity limit of the instrument. Because the number of counts observed when the nova is not in outburst are very small, a background is not removed as part of the aperture photometry. Data from the first, second, fourth and fifth CCD crossings of data are used to estimate the background, which is then fitted with a polynomial, and subtracted from all five years of data. The large scale flatfield corrected and background subtracted light curves over the five year period can be seen in Figure \ref{fig:hi_lc}.

The determination of the photometric response of the HI-2A instrument follows the method of \citet{bewsher10} for the HI-1 instruments. The HI magnitude, $m_{\rm HI}$, scale is given by
\[
  m_{\rm HI}=-2.5\log\frac{I}{F_0},
\]
\noindent where $I$ is the intensity of the star 

and $F_0$ is the intensity of Vega derived from spectral folding (see \citealt{bewsher10} for a detailed discussion) and has a value of 71655.1 DN pixel$^{-1}$. 
\begin{figure}
  \centering
  \includegraphics[width=80mm]{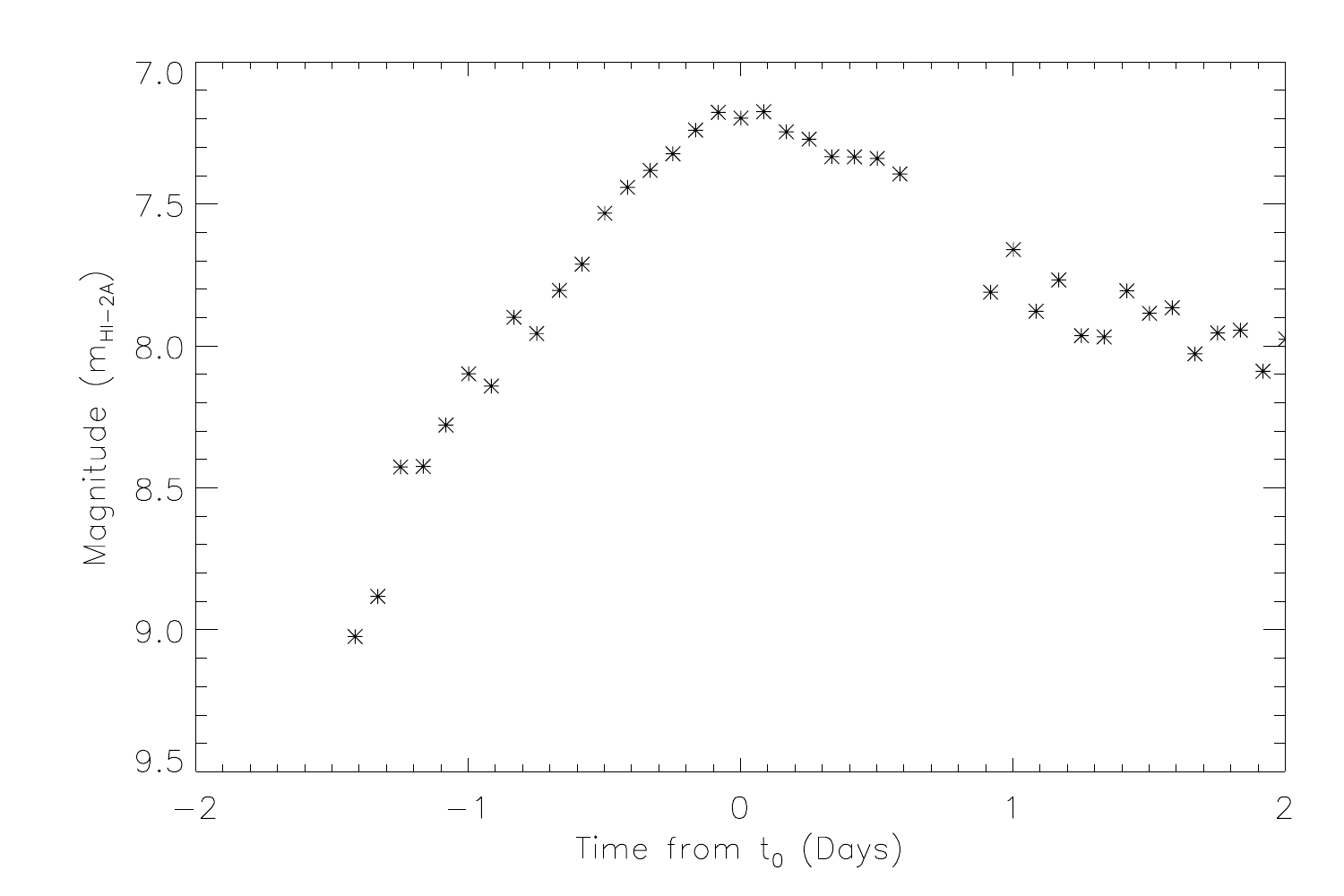}
  \caption{HI magnitude light curve of V5583 Sgr. A close up view of the rise to the time of maximum. The uncertainty is $\sim 0.06$ at magnitudes above 10th magnitude.}
  \label{fig:hi_lc2}
\end{figure}

The optical maxima of novae are usually either missed or poorly constrained in time because of their unexpected nature. However, the 120 minute cadence of the HI-2A camera allows us to place a tight constraint on the time of maximum of V5583~Sgr. The time of maximum ($t_0$) determined from the HI data is JD~$2455050.54$ $\pm0.17$, which translates to 00:09 UT on 2009 August 7th, with an error of $\pm4$\,hours due to the flatness of the peak.  Due to the limiting magnitude of the detector, the rise phase is not seen from the quiescent magnitude ($V>$13.3, \citealt{IAUC9061}). However, \textit{STEREO/HI} did detect the nova $1-2$~days before the optical maximum. We note that the progenitor is not present in the UNSO-B1 catalogue, giving an outburst amplitude of $\Delta B\ge13$, typical of a fast classical nova \citep{bode08}.

Figure \ref{fig:hi_lc2} shows in detail the rise phase of the outburst. It shows a relatively smooth rise, although several data points deviate from the rising curve. The PMHs identified by \citet{hounsell10} develop $0.49-2.19$\,days before maximum, have a duration of $0.14-0.42$\,days, and occur $0.63-1.74~m_{\rm SMEI}$ below maximum, where $m_{\rm SMEI}$ is the unfiltered apparent magnitude derived from the SMEI images. The light curve in Figure~\ref{fig:hi_lc2} shows similar features at 1.2 days before peak, 1.2~$m_{\rm HI}$ below the maximum, 0.9~days before peak, 0.9~$m_{\rm HI}$ below maximum, and 0.8~days before peak, 0.7~$m_{\rm HI}$ below. These are consistent with PMH seen in other novae. Furthermore, \cite{hillman13} have recently observed similar features in their models of nova light curves. They attribute the PMH to the recession of the convection zone and a decline in the opacity of the expanding envelope. Their models show considerable variation in the shape and amplitude of the PMH and some show more than one halt.

The eruption of V5583~Sgr was also detected by the SMEI instrument, but the light curve is noisy because the nova peaked at $m_{\rm SMEI}=6.49\pm0.05$, just one magnitude above limiting value; as a result only a small portion of the rise phase is seen. The time of maximum seen in the SMEI data is  $2455050.6$ (2009 Aug 7.08\,UT). 

\subsection{AAVSO and VSNET Light Curve}
\label{sec:aavso}

\begin{figure}
  \centering
  \includegraphics[trim= 30mm 40mm 10mm 10mm, clip,angle=180,width=80mm]{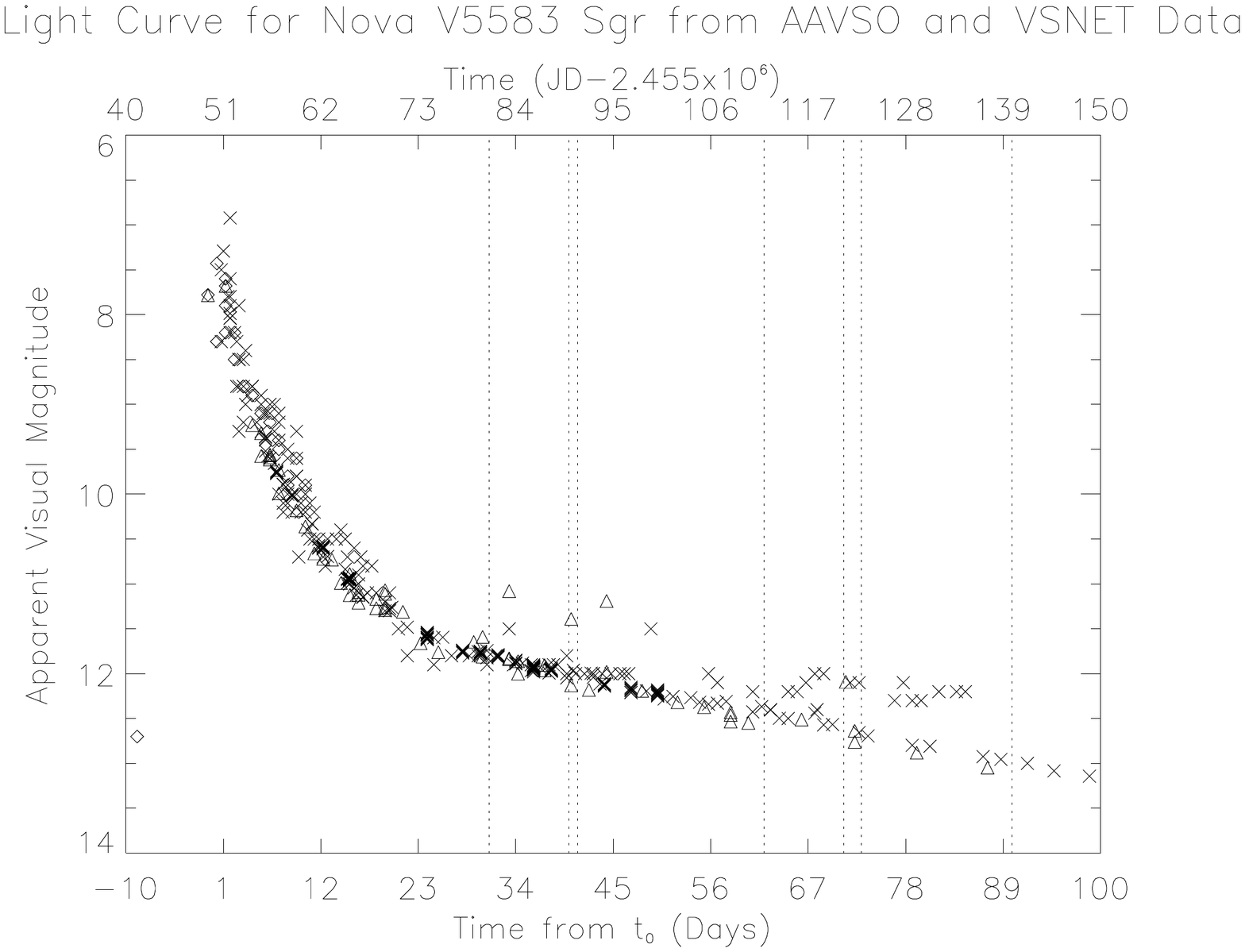}
  
  \caption[Light curve generated from the AAVSO and VSNET data.]{\small{Light curve generated from the AAVSO data (crosses) and VSNET data (open triangles). Additional data points from the IAU circulars are plotted in open diamonds. The vertical lines show the times of the spectroscopic observations discussed in $\S$\ref{sec:optical_spectra}.
}}
  \label{fig:AAVSO_lc}
\end{figure}

The IAUC~9061 observations of V5583~Sgr constrain the time of optical maximum to within approximately 7.6 days, between $29^{th}$ July and 6$^{th}$ August \citep{IAUC9061}. \textit{STEREO/HI} provides a much more precise time, to within just a few hours. The light curve produced from the AAVSO \& VSNET data is shown in Figure \ref{fig:AAVSO_lc}. The first measurement was recorded at JD~$2455051.17$, which translates to 1603 UT on $7^{th}$ August 2009, $\sim22$ hours after the maximum detected in HI. 

The initial decline of the nova, which can be seen in both light curves, is smooth and fast. The AAVSO \& VSNET photometry shows the late decline phase, which is also smooth, implying that there was no dust formation in the ejecta. The behaviour of V5583~Sgr is consistent with a very fast nova \citep{pg57}, with an A type light curve \citep{duerbeck81}. We determine the times to decline by 2 ($t_2$) and 3 ($t_3$) magnitudes from maximum of $t_2=4.97\pm0.32$\,days and $t_3=9.52\pm0.32$\,days from the data in Figure~\ref{fig:AAVSO_lc}; the errors reflect the scatter in the AAVSO \& VSNET light curve ($\sim0.2$ mag) - this will dominate the uncertainty, being more significant than the fading in the visual that occurred in the $\sim22$ hr following the $t_0$ determined from the \textit{STEREO} data. 

Using the MMRD relationship \citep{downes00}, we estimate a magnitude at maximum of $-9.55\pm0.50$ from $t_2$ and $-9.70 \pm1.14$ from $t_3$. The interstellar extinction has been estimated to be $A_V=1.0$ ($E(B-V)=0.33$; \citealt{v5583}), which is consistent with the dust maps of \citet{schlegel98} along the line of sight. These values give an estimate of the distance to the nova of $14.7\pm3.4$\,kpc. 

\section{Optical Spectra}
\subsection{Overview of the spectra}
\label{sec:optical_spectra}

Optical spectra of V5583~Sgr were obtained with the RC Spectrograph, which is mounted on the 1.5m telescope at the Cerro Tololo Inter-American Observatory, Chile. Table~\ref{tab:spectra} logs relevant information on each spectra, see \cite{walter12}.

\begin{table*}
  \centering
  \begin{minipage}{135mm}
    \caption[Log of spectra information.]{Log of spectra information. Classes shown in brackets are alternate classifications, see text for details.}
    \label{tab:spectra}

   \centering
   \begin{tabular}{@{}cccccccc}
     \hline
     
     \multicolumn{1}{c}{\multirow{2}{*}{Date}}&
     \multicolumn{1}{c}{{JD}}&
     \multicolumn{1}{c}{Time since $t_0$}&
     \multicolumn{1}{c}{Slit}&
     \multicolumn{1}{c}{Range}&
     \multicolumn{1}{c}{Resolution}&
     \multicolumn{2}{c}{\multirow{2}{*}{Tololo Class}}\\
     
     \multicolumn{1}{c}{}&
     \multicolumn{1}{c}{(-- 2455000)}&
     \multicolumn{1}{c}{(Days)}&
     \multicolumn{1}{c}{(arcsec)}&
     \multicolumn{1}{c}{(\AA)}&
     \multicolumn{1}{c}{(\AA)}&
     \multicolumn{2}{c}{}\\
     
     \hline
     
     \multicolumn{1}{c}{07/09/2009T03:51}&
     \multicolumn{1}{c}{081}&
     \multicolumn{1}{c}{31}&
     \multicolumn{1}{c}{0.8}&
     \multicolumn{1}{c}{5630-6950}&
     \multicolumn{1}{c}{3.1}&
     \multicolumn{1}{c}{P$_{he}$}&
     \multicolumn{1}{c}{(P$_o^o$)}\\
     
     \multicolumn{1}{c}{16/09/2009T03:27}&
     \multicolumn{1}{c}{090}&
     \multicolumn{1}{c}{40}&
     \multicolumn{1}{c}{1.0}&
     \multicolumn{1}{c}{6000-9500}&
     \multicolumn{1}{c}{8.6}&
     \multicolumn{1}{c}{P$_{o~}^o$}&
     \multicolumn{1}{c}{}\\
     
     \multicolumn{1}{c}{17/09/2009T23:40}&
     \multicolumn{1}{c}{091}&
     \multicolumn{1}{c}{41}&
     \multicolumn{1}{c}{1.0}&
     \multicolumn{1}{c}{3650-5450}&
     \multicolumn{1}{c}{4.1}&
     \multicolumn{1}{c}{N$_{ne}$}&
     \multicolumn{1}{c}{(P$_o^o$)}\\
     
     \multicolumn{1}{c}{08/10/2009T01:48}&
     \multicolumn{1}{c}{112}&
     \multicolumn{1}{c}{62}&
     \multicolumn{1}{c}{1.0}&
     \multicolumn{1}{c}{3650-5450}&
     \multicolumn{1}{c}{4.1}&
     \multicolumn{1}{c}{N$_{ne}$}&
     \multicolumn{1}{c}{}\\
     
     \multicolumn{1}{c}{17/10/2009T00:58}&
     \multicolumn{1}{c}{121}&
     \multicolumn{1}{c}{71}&
     \multicolumn{1}{c}{1.0}&
     \multicolumn{1}{c}{3650-5450}&
     \multicolumn{1}{c}{4.1}&
     \multicolumn{1}{c}{N$_{ne}$}&
     \multicolumn{1}{c}{}\\
     
     \multicolumn{1}{c}{19/10/2009T01:11}&
     \multicolumn{1}{c}{123}&
     \multicolumn{1}{c}{73}&
     \multicolumn{1}{c}{0.8}&
     \multicolumn{1}{c}{5630-6950}&
     \multicolumn{1}{c}{3.1}&
     \multicolumn{1}{c}{P$_{he}$}&
     \multicolumn{1}{c}{(N$_{ne}$)}\\
      
     \multicolumn{1}{c}{05/11/2009T00:23}&
     \multicolumn{1}{c}{140}&
     \multicolumn{1}{c}{90}&
     \multicolumn{1}{c}{1.0}&
     \multicolumn{1}{c}{3650-5450}&
     \multicolumn{1}{c}{4.1}&
     \multicolumn{1}{c}{N$_{ne}$}&
     \multicolumn{1}{c}{}\\

     \hline

    \end{tabular}
  \end{minipage}
\end{table*}

\label{sec:line_id}

\begin{figure}
    \centering
  \includegraphics[trim=62mm 37mm 48mm 21mm,clip,angle=180,width=75mm]{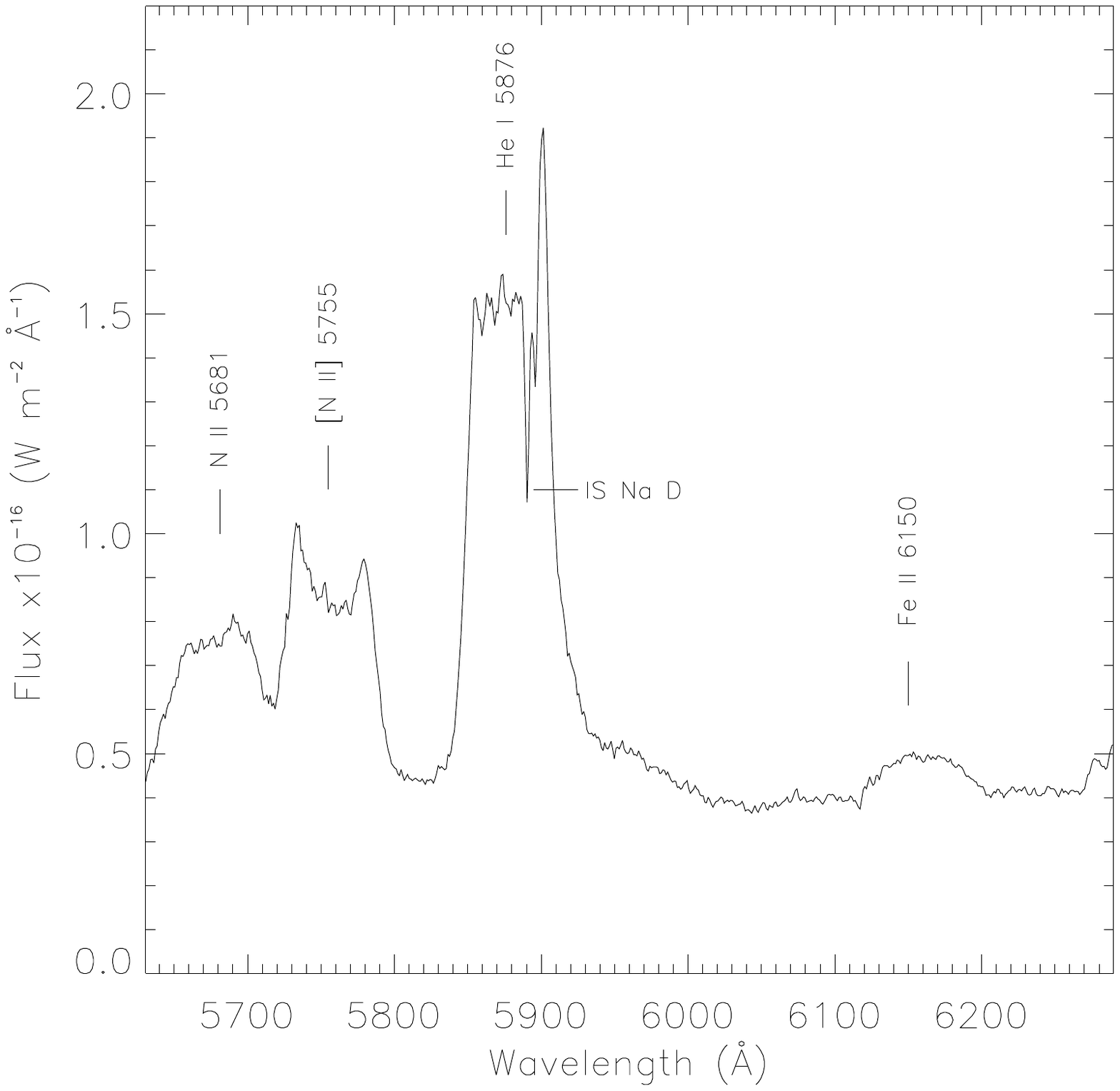}
  \includegraphics[trim=62mm 37mm 48mm 21mm,clip,angle=180,width=75mm]{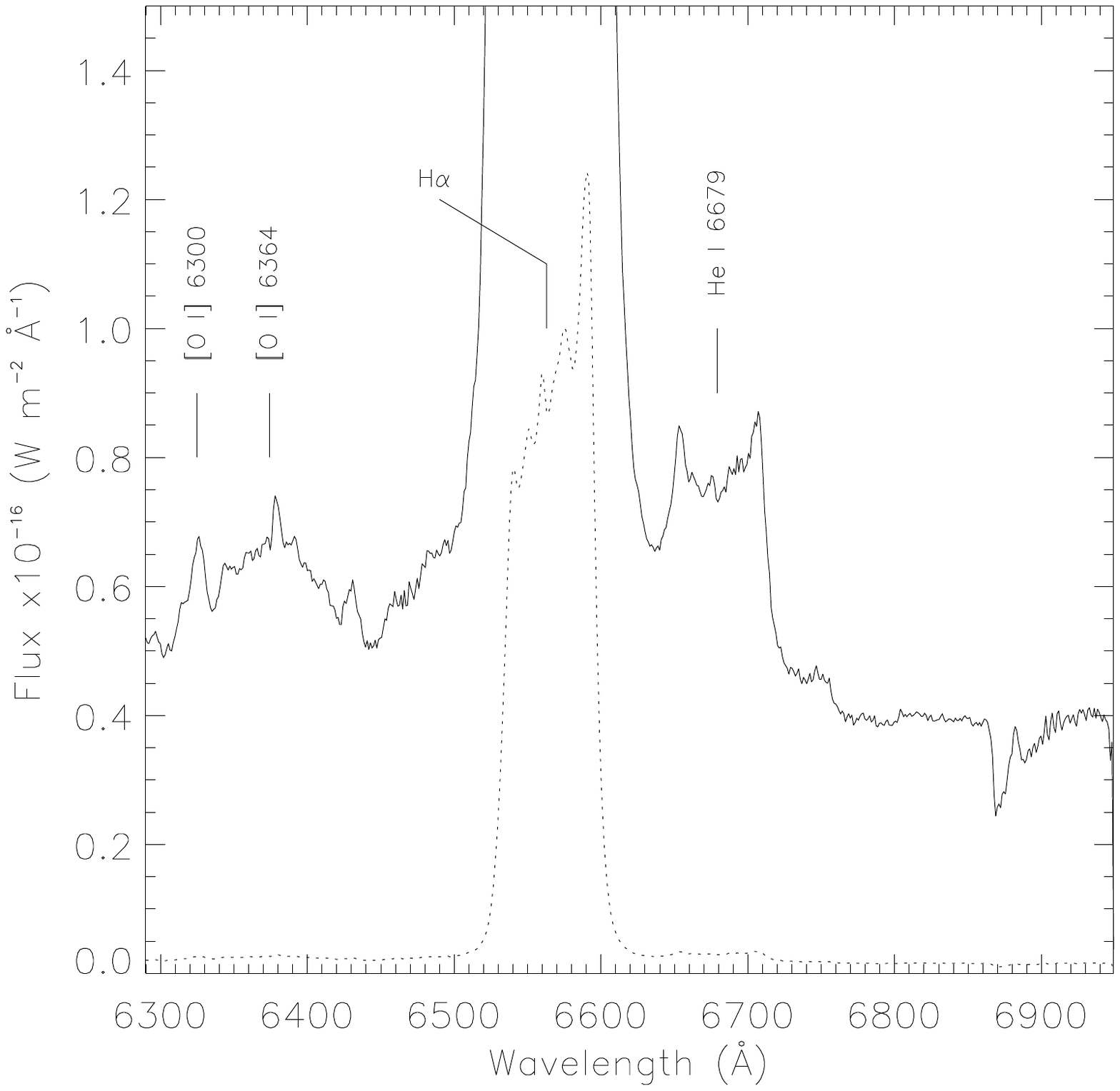}    \caption[Red spectra of V5583 Sgr on 9$^{th}$ September 2009, $t_0+31$.]{Red spectra of V5583 Sgr on 7$^{th}$ September 2009, $t_0+31$. The dotted line in the lower figure shows the profile of H$_\alpha$ at a scale of one quarter.}
  \label{fig:spectra_090907}
\end{figure}

\begin{figure}
    \centering
  \includegraphics[trim=62mm 37mm 48mm 21mm,clip,angle=180,width=75mm]{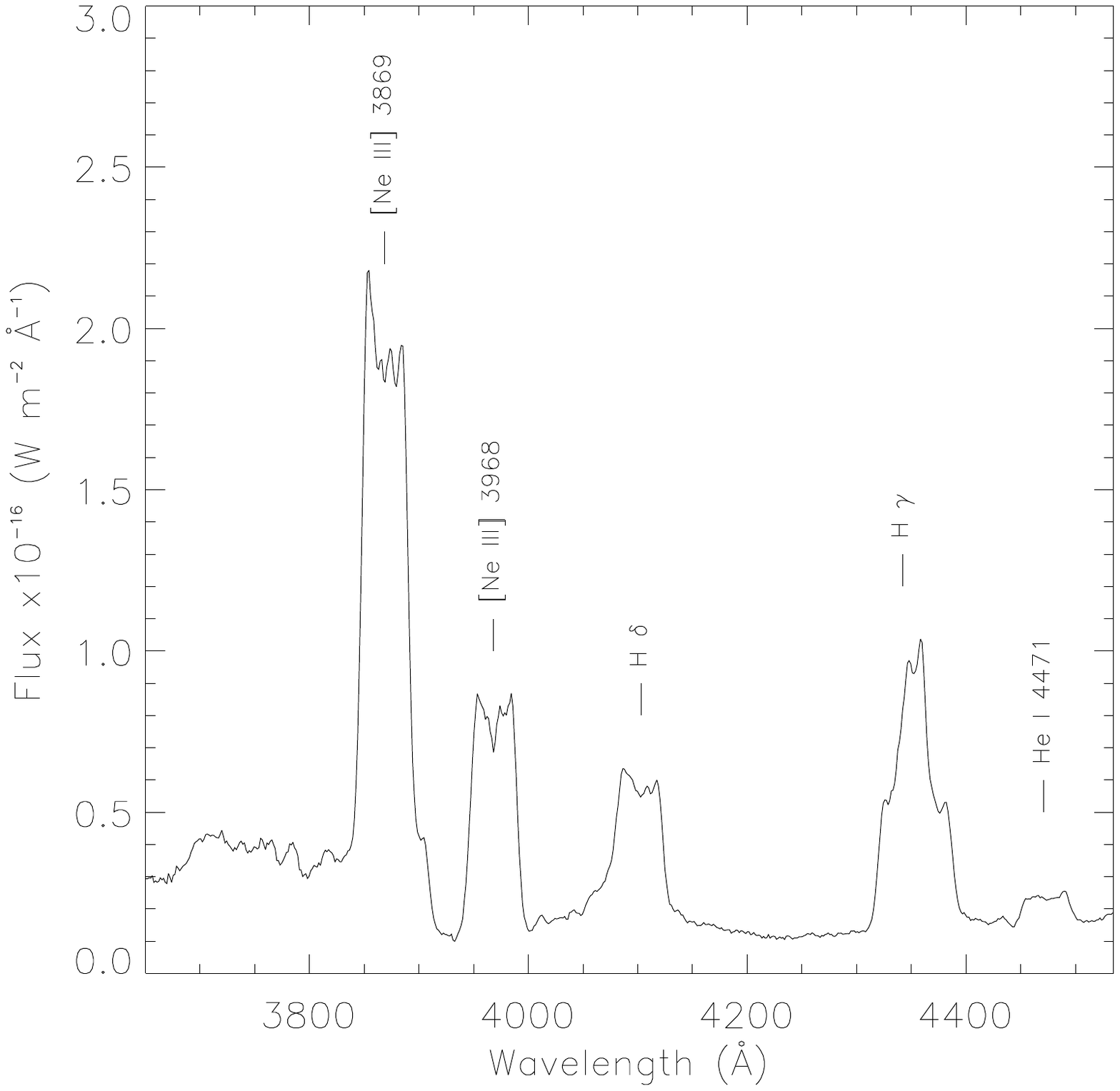}
  \includegraphics[trim=62mm 37mm 48mm 21mm,clip,angle=180,width=75mm]{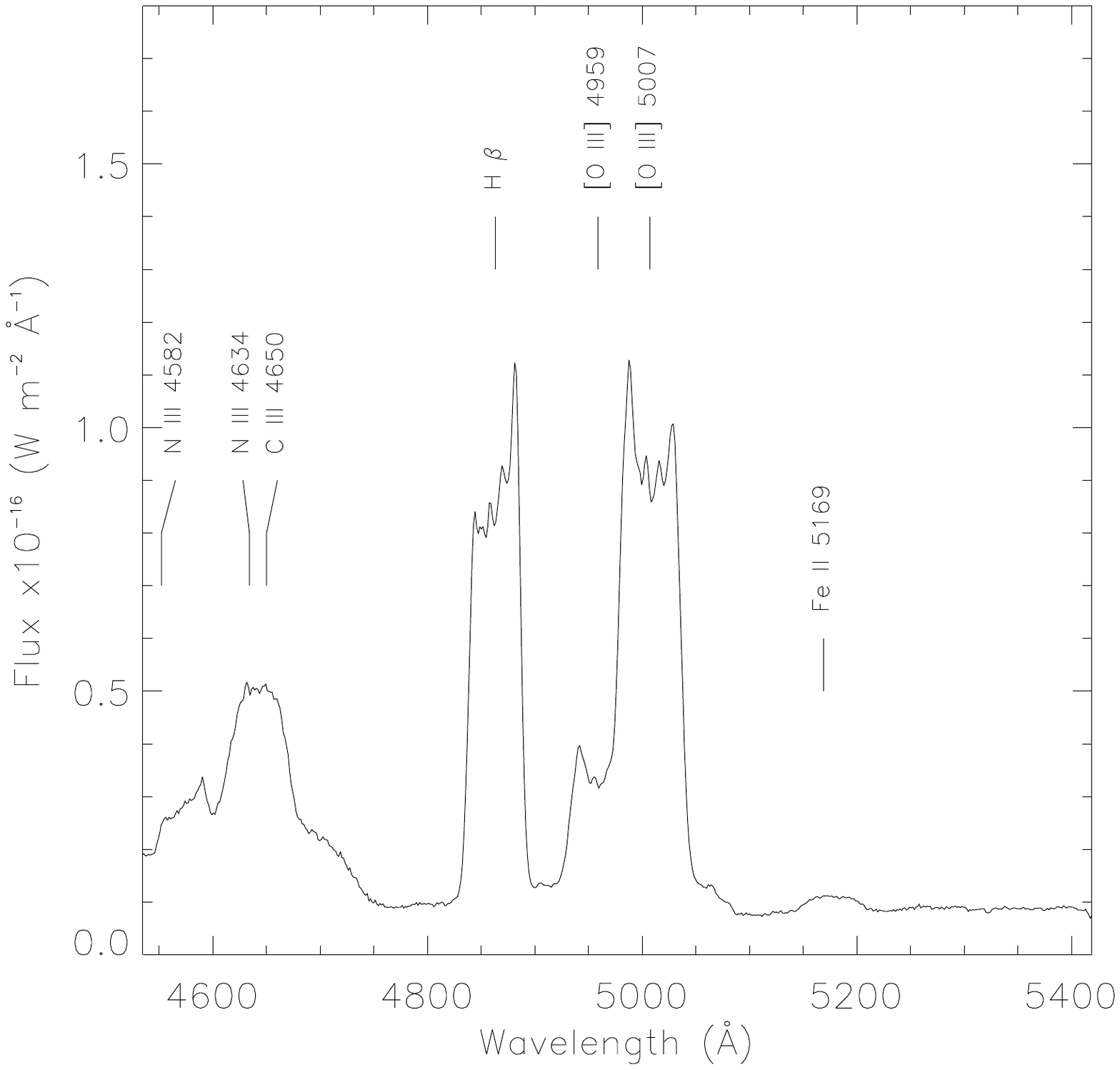}   \caption[Blue spectra of V5583 Sgr on 8$^{th}$ October 2009, $t_0+62$.]{Blue spectra of V5583 Sgr on 8$^{th}$ October 2009, $t_0+62$.}
  \label{fig:spectra_091008}
\end{figure}

Figures \ref{fig:spectra_090907} \& \ref{fig:spectra_091008} show the optical spectra of V5583~Sgr on given dates.
 The spectral lines were identified using an online Atomic Line List (\verb'http://www.pa.uky.edu/~peter/atomic/') and published spectra of fast novae (e.g. \citealt{williams91}, \citealt{augusto03} and \citealt{iijima10}). The times of the spectroscopic observations with respect to the visual light curve of V5583~Sgr are shown in Figure~\ref{fig:AAVSO_lc}. Table~\ref{tab:fluxes} shows the line fluxes on selected dates of observation. The dereddened values have been derived using the extinction law of \citet{seaton79} and a ratio of total-to-selective extinction of $R=3.1$. The smooth light curves are indicative of a dust poor nova, with no significant circumstellar extinction. 

The spectra are dominated by H\,{\sc i}, He\,{\sc i}, O\,{\sc i}, [O\,{\sc iii}], and [Ne\,{\sc iii}] emission lines. The lack of Fe lines suggests that V5583~Sgr was most likely a `He/N' type of classical nova. The later spectra are dominated by the [Ne\,{\sc iii}]$_{3869}$ line, which is second in strength only to H$\alpha$, consistent with a transition from the P$_o^o$ phase to the N$_{ne}$ phase in the Tololo classification scheme of nova spectra. The intermediate auroral phase is not observed in our data, although this phase is by-passed by novae with a ONeMg white dwarf \citep{williams91}. In these novae, the ejecta is enriched in neon, due to mixing of accreted mater with the white dwarf core material \citep{starrfield85}. The strong [Ne\,{\sc iii}] lines in the spectra of V5583~Sgr suggest that there may be an enhancement of this species. 

The line profiles are flat or saddled-shaped and jagged, as is usually the case in `He/N' novae \citep{williams92}. The full width at zero intensity (FWZI) of the lines is 5000\,km\,s$^{-1}$ or greater, consistent with the speed classification derived in $\S$\ref{sec:aavso} \citep{williams92}. The line profiles vary between species and ionisation states.  The [Ne\,{\sc iii}], [N\,{\sc ii}] and [O\,{\sc iii}] lines show peaks at $\sim -1150,~-300,~450,$ and $~1220$\,kms$^{-1}$; in most cases, the blue peak is stronger than the red peak. The He\,{\sc i}, H\,{\sc i} and O\,{\sc i} also show these velocities, but the red peak is stronger than the blue peak. These line profiles are often seen in novae and are attributed to inhomogenities and  clumpiness in the ejecta \citep{gill}.

\subsection{The neon abundance}

\begin{figure}
  \centering
  \includegraphics[trim= 20mm 130mm 20mm 30mm, clip,angle=0,width=80mm]{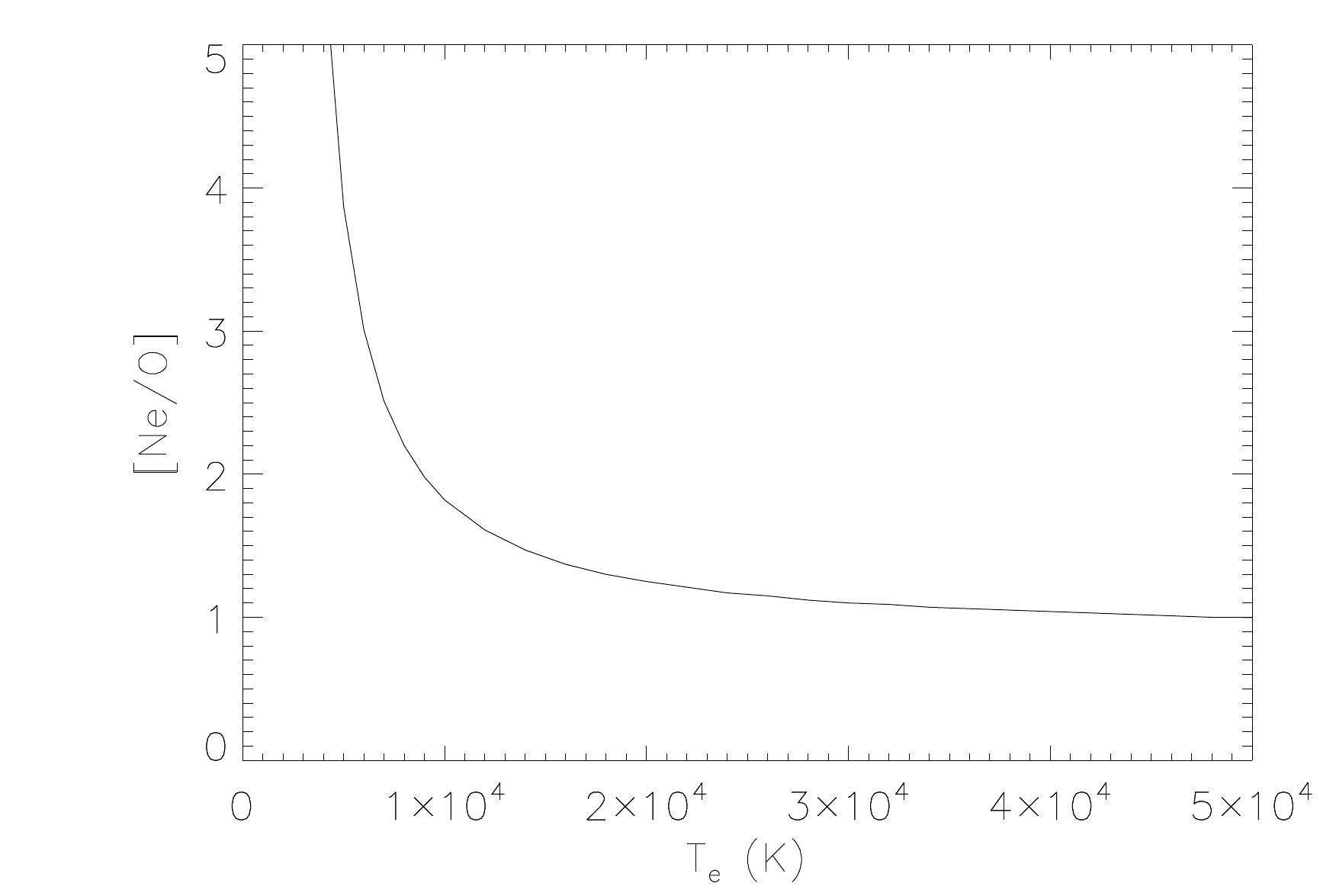}
  
  \caption{The [Ne/O] abundance in V5583~Sgr as a function of assumed electron temperature.
}
  \label{neon}
\end{figure}

To determine the abundance of neon in the ejecta of V5583~Sgr, we consider a line ratio from the species  [Ne\,{\sc iii}] and [O\,{\sc iii}], which have similar ionisation potentials (41\,eV and 35\,eV respectively). If the density is $n_e>10^5$\,cm$^{-3}$, the line flux ($F_{ij}$) is given by the formula:
\[
  F_{ij} = N_i E_{ij}A_{ij}\frac{g_j}{g_i}\exp{\frac{-E_{ij}}{kT_e}},
\]
where $N_i$ is the number density of level $i$, $E_{ij}$ is the energy of the transition, $A_{ij}$ is the transition probability, $g$ is the statistical weight, and $T_e$ is the electron temperature.
In principal, $T_e$ can be determined from the flux ratios [O\,{\sc iii}] $(4959+5007)/4363$ and [N\,{\sc ii}] $(6548+6584)/5755$ \citep{osterbrock}, 
but these are severely effected by blending with H\,{\sc i} lines in the data. 
However, in Figure~\ref{neon} we show the [Ne/O] abundance in V5583~Sgr for a range of electron temperatures from which we can make a meaningful constraint. 
We note that the H$\beta$/H$\gamma$ line ratio is lower than expected, and that the [O\,{\sc iii}] 4363 line must contribute significantly to the H$\gamma$ flux; this shows that collisional effects are affecting the population of the [O\,{\sc iii}] levels, which suggests $n_e>10^5$cm$^{-3}$ \citep{osterbrock}. Furthermore, we extract the [O\,{\sc iii}] 4363\,\AA\, line flux from the blend with H$\gamma$ using the H$\beta$ profile. This gives the flux ratio [O\,{\sc iii}] $(4959+5007)/4363=2.92$. The density is about $1.5\times10^7$\,cm$^{-3}$ at $T_{\rm e}=10,000$\,K, $2\times10^6$\,cm$^{-3}$ at 50,000K and $4\times10^6$\,cm$^{-3}$ at 10$^6$\,K. 

We consider the data on days 62, 71 and 90, in which the relevant lines are relatively free of contamination. Although [O\,{\sc iii}] 5007 forms a blend with [O\,{\sc iii}] 4959, we can extract its flux using the transition probabilities, since these lines share the same upper level; this gives a ratio of $5007/4959=2.9$.
In deriving the [Ne/O] abundance, we average the results from each dataset, which give similar values. We give the results with respect to solar, assuming the abundances in \cite{asplund}.

Figure~\ref{neon} shows that we would find lower [Ne/O] abundances if we assume higher assumed electron temperatures, with the abundance asymptotically approaching a value of about $\rm [Ne/O]\ga+1.0$ for $T_e>20\times10^4$\,K. We can therefore conclude that [Ne/O] is well in excess of the solar value in the ejecta of V5583~Sgr. Such a high abundance is usually attributed to a nova occurring on a ONeMg WD, in which some of the WD material has been dredged-up and ejected in the eruption, while novae occurring on CO WDs have $\rm [Ne/O]<+0.11$ \citep{Gehrz08}. It is worth noting that $\rm [Ne/O]>+1.0$ places V5583~Sgr among the extreme neon novae, along with V693~CrA and V1370~Aql \citep{Andrea94}. 

\begin{table*}
  \centering
  \begin{minipage}{135mm}
    \caption{Table of line fluxes on selected dates. Notes: 1) $^{\diamond}$ indicates a line with an absorption component.}
    \label{tab:fluxes}
    \centering
    \begin{tabular}{cccccc}

      \hline
      \multicolumn{1}{c}{\multirow{3}{*}{Line}}&
      \multicolumn{1}{c}{\multirow{2}{*}{Time since $t_0$}}&
      \multicolumn{1}{c}{\multirow{2}{*}{JD}}&
      \multicolumn{1}{c}{Flux}&
      \multicolumn{1}{c}{Extinction corrected}&
      \multicolumn{1}{c}{FWZI}\\
      
      \multicolumn{1}{c}{}&
      \multicolumn{1}{c}{\multirow{2}{*}{(Days)}}&
      \multicolumn{1}{c}{\multirow{2}{*}{(-- 2455000)}}&
      \multicolumn{1}{c}{$\times 10^{-15}$}&
      \multicolumn{1}{c}{(Flux) $\times 10^{-15}$}&
      \multicolumn{1}{c}{$\pm500$}\\
       
      \multicolumn{1}{c}{}&
      \multicolumn{1}{c}{}&
      \multicolumn{1}{c}{}&
      \multicolumn{1}{c}{(W m$^{-2}$)}&
      \multicolumn{1}{c}{(W m$^{-2}$)}&
      \multicolumn{1}{c}{(km s$^{-1}$)}\\

      \hline
      
      \multicolumn{1}{c}{H$_\delta$}&
      \multicolumn{1}{c}{62}&
      \multicolumn{1}{c}{112}&
      \multicolumn{1}{c}{$2.43\pm0.13$}&
      \multicolumn{1}{c}{$8.89\pm2.05$}&
      \multicolumn{1}{c}{7000}\\
  
      \multicolumn{6}{c}{}\\
      
      \multicolumn{1}{c}{H$_\gamma$}&
      \multicolumn{1}{c}{62}&
      \multicolumn{1}{c}{112}&
      \multicolumn{1}{c}{$4.70\pm0.12$}&
      \multicolumn{1}{c}{$16.6\pm3.8$}&
      \multicolumn{1}{c}{11000}\\
       
      \multicolumn{6}{c}{}\\

      \multicolumn{1}{c}{H$_\beta$}&
      \multicolumn{1}{c}{62}&
      \multicolumn{1}{c}{112}&
      \multicolumn{1}{c}{$4.10\pm0.21$}&
      \multicolumn{1}{c}{$12.9\pm3.0$}&
      \multicolumn{1}{c}{6000}\\
       
      \multicolumn{6}{c}{}\\
      
      \multicolumn{1}{c}{H$_\alpha$}&
      \multicolumn{1}{c}{40}&
      \multicolumn{1}{c}{90}&
      \multicolumn{1}{c}{$114\pm12$}&
      \multicolumn{1}{c}{$240\pm55$}&
      \multicolumn{1}{c}{6000}\\
      
      \multicolumn{6}{c}{}\\
      
      \multicolumn{1}{c}{H \sc{i}$_{8865}$}&
      \multicolumn{1}{c}{40}&
      \multicolumn{1}{c}{90}&
      \multicolumn{1}{c}{$1.40\pm0.12$}&
      \multicolumn{1}{c}{$2.18\pm0.50$}&
      \multicolumn{1}{c}{5000}\\
        
      \multicolumn{6}{c}{}\\

      \multicolumn{1}{c}{H {\sc{i}}$_{9017}$}&
      \multicolumn{1}{c}{40}&
      \multicolumn{1}{c}{90}&
      \multicolumn{1}{c}{$1.89\pm0.28$}&
      \multicolumn{1}{c}{$2.93\pm0.67$}&
      \multicolumn{1}{c}{4700}\\
       
      \multicolumn{6}{c}{}\\
      
      \multicolumn{1}{c}{H \sc{i}$_{9231}$}&
      \multicolumn{1}{c}{40}&
      \multicolumn{1}{c}{90}&
      \multicolumn{1}{c}{$5.20\pm0.39$}&
      \multicolumn{1}{c}{$8.06\pm1.85$}&
      \multicolumn{1}{c}{8300}\\
        
      \multicolumn{6}{c}{}\\
      
      \multicolumn{1}{c}{He \sc{i}$_{4471}$}&
      \multicolumn{1}{c}{62}&
      \multicolumn{1}{c}{112}&
      \multicolumn{1}{c}{$0.93\pm0.12$}&
      \multicolumn{1}{c}{$3.11\pm0.72$}&
      \multicolumn{1}{c}{11200}\\
 
      \multicolumn{6}{c}{}\\
      
      \multicolumn{1}{c}{He \sc{i}$_{5876}$}&
      \multicolumn{1}{c}{73$^{\diamond}$}&
      \multicolumn{1}{c}{123}&
      \multicolumn{1}{c}{$3.35\pm0.27$}&
      \multicolumn{1}{c}{$8.04\pm1.85$}&
      \multicolumn{1}{c}{5600}\\
  
      \multicolumn{6}{c}{}\\
      
      \multicolumn{1}{c}{He \sc{i}$_{6679}$}&
      \multicolumn{1}{c}{40}&
      \multicolumn{1}{c}{90}&
      \multicolumn{1}{c}{$2.74\pm0.06$}&
      \multicolumn{1}{c}{$5.74\pm1.32$}&
      \multicolumn{1}{c}{6600}\\
       
      \multicolumn{6}{c}{}\\
      
      \multicolumn{1}{c}{He \sc{i}$_{7065}$}&
      \multicolumn{1}{c}{40}&
      \multicolumn{1}{c}{90}&
      \multicolumn{1}{c}{$6.67\pm0.19$}&
      \multicolumn{1}{c}{$13.2\pm3.0$}&
      \multicolumn{1}{c}{6800}\\
       
      \multicolumn{6}{c}{}\\
      
      \multicolumn{1}{c}{C \sc{iii}$_{8197}$}&
      \multicolumn{1}{c}{40}&
      \multicolumn{1}{c}{90}&
      \multicolumn{1}{c}{$0.71\pm0.02$}&
      \multicolumn{1}{c}{$1.24\pm0.29$}&
      \multicolumn{1}{c}{4800}\\
       
      \multicolumn{6}{c}{}\\
      
      \multicolumn{1}{c}{N \sc{iii}$_{4582}$}&
      \multicolumn{1}{c}{62}&
      \multicolumn{1}{c}{112}&
      \multicolumn{1}{c}{$1.66\pm0.20$}&
      \multicolumn{1}{c}{$5.54\pm1.28$}&
      \multicolumn{1}{c}{16500}\\

      \multicolumn{6}{c}{}\\
      
      \multicolumn{1}{c}{N \sc {ii}$_{5681}$}&
      \multicolumn{1}{c}{73}&
      \multicolumn{1}{c}{123}&
      \multicolumn{1}{c}{$0.51\pm0.05$}&
      \multicolumn{1}{c}{$1.32\pm0.30$}&
      \multicolumn{1}{c}{5818}\\
       
      \multicolumn{6}{c}{}\\
      
      \multicolumn{1}{c}{[N \sc {ii}]$_{5755}$}&
      \multicolumn{1}{c}{73}&
      \multicolumn{1}{c}{123}&
      \multicolumn{1}{c}{$0.85\pm0.422$}&
      \multicolumn{1}{c}{$2.04\pm0.47$}&
      \multicolumn{1}{c}{5400}\\
       
      \multicolumn{6}{c}{}\\

      \multicolumn{1}{c}{[O \sc {iii}]$_{4959}$}&
      \multicolumn{1}{c}{62}&
      \multicolumn{1}{c}{112}&
      \multicolumn{1}{c}{$1.22\pm0.09$}&
      \multicolumn{1}{c}{$3.61\pm0.83$}&
      \multicolumn{1}{c}{5100}\\
       
      \multicolumn{6}{c}{}\\
      
      \multicolumn{1}{c}{[O \sc {iii}]$_{5007}$}&
      \multicolumn{1}{c}{62}&
      \multicolumn{1}{c}{112}&
      \multicolumn{1}{c}{$5.48\pm0.33$}&
      \multicolumn{1}{c}{$16.2\pm3.7$}&
      \multicolumn{1}{c}{7300}\\
      
      \multicolumn{6}{c}{}\\

      \multicolumn{1}{c}{[O \sc{iii}]$_{7325}$}&
      \multicolumn{1}{c}{40}&
      \multicolumn{1}{c}{90}&
      \multicolumn{1}{c}{$6.27\pm1.00$}&
      \multicolumn{1}{c}{$12.4\pm2.9$}&
      \multicolumn{1}{c}{7100}\\
         
      \multicolumn{6}{c}{}\\
      
      \multicolumn{1}{c}{O \sc{i}$_{7776}$}&
      \multicolumn{1}{c}{40}&
      \multicolumn{1}{c}{90}&
      \multicolumn{1}{c}{$2.03\pm0.08$}&
      \multicolumn{1}{c}{$3.78\pm0.87$}&
      \multicolumn{1}{c}{5100}\\
        
      \multicolumn{6}{c}{}\\
      
      \multicolumn{1}{c}{O \sc {i}$_{8221}$}&
      \multicolumn{1}{c}{40}&
      \multicolumn{1}{c}{90}&
      \multicolumn{1}{c}{$19.5\pm0.7$}&
      \multicolumn{1}{c}{$34.2\pm7.9$}&
      \multicolumn{1}{c}{4600}\\

      \multicolumn{6}{c}{}\\
      
      \multicolumn{1}{c}{O \sc {i}$_{8446}$}&
      \multicolumn{1}{c}{40}&
      \multicolumn{1}{c}{90}&
      \multicolumn{1}{c}{$29.8\pm4.3$}&
      \multicolumn{1}{c}{$52.1\pm12.0$}&
      \multicolumn{1}{c}{4000}\\
      
      \multicolumn{6}{c}{}\\

      \multicolumn{1}{c}{[Ne \sc {iii}]$_{3869}$}&
      \multicolumn{1}{c}{62}&
      \multicolumn{1}{c}{112}&
      \multicolumn{1}{c}{$8.22\pm0.34$}&
      \multicolumn{1}{c}{$32.5\pm7.5$}&
      \multicolumn{1}{c}{8600}\\
       
      \multicolumn{6}{c}{}\\
      
      \multicolumn{1}{c}{[Ne \sc {iii}]$_{3968}$}&

      \multicolumn{1}{c}{62}&
      \multicolumn{1}{c}{112}&
      \multicolumn{1}{c}{$3.26\pm0.23$}&
      \multicolumn{1}{c}{$12.4\pm2.9$}&
      \multicolumn{1}{c}{5200}\\
      
      \multicolumn{6}{c}{}\\
      
      \multicolumn{1}{c}{Ne \sc {ii}$_{6168}$}&
      \multicolumn{1}{c}{40}&
      \multicolumn{1}{c}{90}&
      \multicolumn{1}{c}{$0.73\pm0.02$}&
      \multicolumn{1}{c}{$1.65\pm0.38$}&
      \multicolumn{1}{c}{10900}\\
       
      \multicolumn{6}{c}{}\\
    
      \multicolumn{1}{c}{Fe \sc {ii}$_{5169}$}&
      \multicolumn{1}{c}{62}&
      \multicolumn{1}{c}{112}&
      \multicolumn{1}{c}{$0.20\pm0.03$}&
      \multicolumn{1}{c}{$0.548\pm0.126$}&
      \multicolumn{1}{c}{7000}\\

      \hline
      
    \end{tabular}
  \end{minipage}
\end{table*}

\section{Discussion}
\label{sec:opt_discussion}

In this paper we have presented a light curve of the very fast classical nova V5583~Sgr, obtained by the HI-2A camera on board the \textit{STEREO} spacecraft.  The extracted data show a clear detection of the nova outburst, and include the rarely seen phase before maximum.

The cadence of the HI-2 instrument is 120 minutes, giving a tight constraint on the time of the maximum light. The derived $t_0$ is JD~$2455050.54\pm 0.17$,  two days earlier than that estimated by \citet{v5583}, who assumed that V5583~Sgr was identical to another nova, V382~Vel. In the HI light curve of V5583~Sgr, a PMH may have occurred one day before maximum (as shown in Figure \ref{fig:hi_lc2}), similar to those of RS~Oph and KT~Eri in the SMEI data \citep{hounsell10}. 

The detection of PMHs is rare due to the unpredictable nature of novae. However, it is possible that other novae are present in the HI archive data. Furthermore, any nova detected with the HI-1 instruments, with their smaller pixel size and higher cadence, will be shown in unprecedented detail.
 
The optical spectra of V5583~Sgr show an emission line spectrum typical of a fast nova in the decline stage. Most of the spectra have been classified as N$_{ne}$ in the Tololo system, with the remaining being P$^o_o$. None are consistent with the coronal or auroral phase, which suggests that V5583~Sgr is a neon nova \citep{williams92}. The profiles of the spectral lines, and the FWZIs are also consistent with this interpretation, as is our conclusion that the object is a `He/N' type nova \citep{williams92}. 

We have derived a neon abundance in the ejecta of $\rm [Ne/O]\ga+1.0$, which shows that V5583~Sgr was likely to be a neon nova, and that material has been dredged-up from the white dwarf and ejected in the explosion. We can therefore conclude that the primary is a massive ($M_{WD}\ga1.2M_{\sun}$) ONeMg white dwarf \citep{Gehrz08}, which is also consistent with the speed classification.

V382~Vel, which was discovered in outburst in 1999, is, in many respects, similar to V5583~Sgr. It, too, was a very fast, neon-type nova, with a neon abundance of $\rm [Ne/O]=+0.66$ and $t_2$ of 4.5 days, and showed little or no evidence of dust formation in the ejecta \citep{dv02}. Early spectroscopic observations ($t_0+5$) showed many Fe {\sc{ii}} lines in its spectra \citep{dv02} which disappeared about one month later. The spectra following this epoch, when the nova was in the nebular phase, are almost identical to those presented here for V5583~Sgr over the same evolutionary period. It is possible that earlier spectra of V5583~Sgr would have also shown Fe~{\sc{ii}} lines, similar to those seen in V382~Vel. The latest observations of V382~Vel, $\sim184$ days after maximum, showed highly ionised Fe lines, which are not present in the V5583~Sgr data. From these spectra, \citet{dv02} classify the object as an `Fe {\sc{ii}}b' type nova. On the basis of the lines present in our spectra of V5583~Sgr, we classified the object as a `He/N' type nova; however, given the similarity to V382~Vel, it is possible that V5583~Sgr is a `Fe {\sc{ii}}b' nova, and would have developed Fe lines at a later stage. 

\section{Conclusions}

In this paper we have presented \textit{STEREO/HI} observations of V5583~Sgr, which tightly constrains the time of maximum light at JD~$2455050.54\pm0.17$. The $t_2$ of the nova is $4.97\pm0.32$ days, making V5583~Sgr a very fast nova.

Optical spectroscopy of V5583~Sgr show the nova evolving from the permitted phase to the nebula phase.  The neon abundance in the ejecta is $\rm [Ne/O]\ga+1.0$, which makes V5583~Sgr one of the most neon enriched novae known.

\section*{Acknowledgements}

We acknowledge with thanks the variable star observations from the AAVSO and VSNET international databases contributed by observers worldwide and used in this research. 
The Heliospheric Imager (HI) instrument was developed by a collaboration
that included the Rutherford Appleton Laboratory and the University of
Birmingham, both in the United Kingdom, and the Centre Spatial de Li\`{e}ge
(CSL), Belgium, and the US Naval Research Laboratory (NRL),Washington DC,
USA. The STEREO/SECCHI project is an international consortium of the Naval
Research Laboratory (USA), Lockheed Martin Solar and Astrophysics Lab
(USA), NASA Goddard Space Flight Center (USA), Rutherford Appleton
Laboratory (UK), University of Birmingham (UK), Max-Planck-Institut f\"{u}r
Sonnen- systemforschung (Germany), Centre Spatial de Li\`{e}ge (Belgium),
Institut d$'$Optique Th\'{e}orique et Appliqu\'{e}e (France), and Institut
d$'$Astrophysique Spatiale (France).This research has made use of the SIMBAD database, operated at CDS, Strasbourg, France. The authors thank the anonymous reviewer for useful comments.

\bibliographystyle{mn2e}
\bibliography{DH_V5583Sgr}

\label{lastpage}
\end{document}